# Origin of Oxygen Partial Pressure-Dependent Conductivity in SrTiO$_3$

Zenghua Cai[1,2,a)] and Chunlan Ma[1,2]

[1]Key Laboratory of Intelligent Optoelectronic Devices and Chips of Jiangsu Higher Education Institutions, School of Physical Science and Technology, Suzhou University of Science and Technology, Suzhou, 215009, China
[2]Advanced Technology Research Institute of Taihu Photon Center, School of Physical Science and Technology, Suzhou University of Science and Technology, Suzhou, 215009, China

[a)]**Authors to whom correspondence should be addressed:** zhcai@usts.edu.cn

**Abstract**

SrTiO$_3$ (STO) displays a broad spectrum of physical properties, including superconductivity, ferroelectricity, and photoconductivity, making it a standout semiconductor material. Despite extensive researches, the oxygen partial pressure-dependent conductivity in STO has remained elusive. This study leverages first-principles calculations, and systematically investigates the intrinsic defect properties of STO. The results reveal that $V_O$, $V_{Sr}$, and $Ti_{Sr}$ are the dominant intrinsic defects, influencing STO's conductivity under varying O chemical potentials (oxygen partial pressures). Under O-poor condition, $V_O$ is the predominant donor, while $V_{Sr}$ is the main acceptor. As the oxygen pressure increases, $Ti_{Sr}$ emerges as a critical donor defect under O-rich condition, significantly affecting the conductivity. Additionally, the study elucidates the abnormal phenomenon where $V_{Ti}$, typically an acceptor, exhibits donor-like behavior due to the formation of O-trimer. This work offers a comprehensive understanding of how intrinsic defects tune the Fermi level, thereby altering STO's conductivity from metallic to n-type, and eventually to p-type across different O chemical potentials. These insights resolve the long-standing issue of oxygen partial pressure-dependent conductivity and explain the observed metallic conductivity in oxygen-deficient STO.

As a prototypical perovskite, SrTiO$_3$ (STO) possesses many fascinating physical properties such as superconductivity[1], ferroelectricity[2-4], photoconductivity[5], two-dimensional (2D) electron gas[6], blue-light emission[7], magnetic effect[8], thermoelectric coefficient[9], metal-insulator transition[10], spin-charge conversion[11], and spin splitting[12]. Recently, two new physical properties have also been reported, i.e., terahertz electric-field-driven dynamical multiferroicity[13] and highly reversible extrinsic electrocaloric effect[14]. With all these attractive properties, STO has become a unique semiconductor which has received much attention in recent decades.

Point defects are of special importance to semiconductors. As a unique semiconductor, the defect properties of STO are undoubtedly important and have been intensively studied. In experiment, studies on the defect properties of STO can mainly be divided into two aspects: (i) focusing on the defect itself, such as the identification of cation vacancies ($V_{Ti}$ and $V_{Sr}$)[15], direct observation of Sr vacancy ($V_{Sr}$)[16], discovering the abundant Ti$_{Sr}$ antisite[17], and confirming the existence of Ti vacancy ($V_{Ti}$)[18]; (ii) the effects induced by the defect, such as high electron mobility induced by the $V_{Sr}$ clusters[19], photo-activated electron transport governed by defect complexes[20], photoconductivity related to $V_{Sr}$[21], and photoflexoelectricity enhanced by the oxygen vacancy ($V_O$)[22]. Similarly, theoretical investigations of defects in STO can also be classified into the same two aspects. For the defect itself, the studies includes intrinsic defects[23-28], impurity defects[29-32], and different methods for simulating defect properties in STO[33-37]. For the defect correlated effects, studies include the coloration in Fe-doped STO[38], high temperature conductivity influenced by impurity defects[39], persistent photoconductivity caused by substitutional hydrogen ($H_O$)[40], ferroelectricity and blue light emission related to Ti antisite defect ($Ti_{Sr}$)[41], and so on[42-45].

Even though so much effort has been made, a long-standing issue correlated with defects, i.e., the origin of oxygen partial pressure-dependent conductivity, still remains unsolved. As early as 1995, Akhtar et al. has pointed out this issue[46]. As the increase of the oxygen partial pressure, the conductivity of STO first decreases and then increases. Meanwhile, the type of conductivity changes from n-type to p-type. In 2003, Tanaka et al. explained this issue from the perspective of intrinsic vacancies based on local density approximation (LDA)[47]. More recently, Wu et al. performed a grand canonical multiscale simulation and analyzed this issue from the perspective of grain size[48]. As is well known, the conductivity of a semiconductor (without doping) is mainly determined by the intrinsic point defects, including vacancy, antisite and interstitial. However, a systematic study on the oxygen partial pressure-dependent conductivity in STO from the perspective of intrinsic point defects is still lacking. Hence, we performed a systematic first-principles study on the intrinsic point defect properties of STO in this work.

Defect properties are mainly determined by the structures of defects. In order to predict the defect properties precisely, it is necessary to locate the ground-state structures accurately. Therefore, we used the ShakeNBreak to find out the ground-state structures of defects in this work[49]. Since it is very time-consuming to relax different distorted defect structures, the bond distortions are only considered for neutral defects. For example, Fig. 1(a) shows the final total energies of neutral $V_O$ versus bond distortion factor (see supplementary materials for computational details). As we can see, 9 distorted structures are considered for neutral $V_O$. Most of the distorted structures have the similar total energies around -1069 eV after relaxation (for 135-atom supercell). Only one structure has a relatively high total energy around -1062 eV, which is confirmed unreasonable after checking the relaxed structure. Among the remaining 8 distorted structures, they all have similar configurations after relaxation as shown in Fig. 1(b). Two Ti atoms (adjacent to $V_O$) move slightly away from the position of $V_O$, consistent with the previous reports[25,36]. The final ground-state structure for neutral $V_O$ is taken as the one without initial distortions, marked with a red star in Fig. 1(a), since this structure has the lowest total energy. Moreover, for the charged defects, their initial structures (before relaxation) come from the related neutral ground-state structures identified by ShakeNBreak.

In order to find out the dominant intrinsic defects affecting the conductivity, the formation energies of all the non-equivalent intrinsic defects are calculated, including vacancies ($V_{Sr}$, $V_{Ti}$, $V_O$), antisites ($Sr_{Ti}$, $Sr_O$, $Ti_{Sr}$, $Ti_O$, $O_{Sr}$, $O_{Ti}$) and interstitials. For the interstitials, five random non-equivalent interstitial sites are considered for Sr, Ti and O atoms. Then, the lowest energy configuration is picked as the candidate. For example, the second interstitial site for Sr ($Sr_{i2}$) is the most stable, hence $Sr_{i2}$ is selected as the candidate for Sr interstitial. For Ti and O atoms, $Ti_{i3}$ and $O_{i1}$ are chosen to represent the Ti and O interstitials, respectively. Fig. 2 shows the calculated formation energies as function of Fermi level ($E_F$) under different O chemical potentials (see supplementary materials for the selection of chemical potential). Under O-poor condition (low oxygen partial pressure), the dominant donor defect is $V_O$ and it is always in +2 charge state. This means there is no defect transition level for $V_O$ in the band gap as shown in Fig. 3(d). Meanwhile, the dominant acceptor defect is $V_{Sr}$. It has a very shallow transition level from -1 to -2 charge state, and this level is 0.05 eV above the valence band maximum (VBM). Moreover, the donor defects $Ti_{Sr}$, $Ti_{i3}$, $Ti_O$ and $Sr_O$ also have relatively low formation energies, while according to the analysis of defect densities as shown in Fig. 4, the densities of these donor defects are relatively low. Hence, their influence on the conductivity is limited, and they are not the dominant donor defects under O-poor condition.

As the oxygen partial pressure increases, the chemical potential of O will change to mediate condition. In this situation, the prominent donor and acceptor defects are still $V_O$ and $V_{Sr}$ as shown in Fig. 2(b). As the oxygen partial pressure increases further, the chemical potential of O will become rich. At this moment, the dominant acceptor defect is still $V_{Sr}$, while the dominant donor defects become $V_O$ and $Ti_{Sr}$. Although the formation energy of $Ti_{Sr}$ has an upward trend with the increase of the O chemical potential, its contribution to conductivity increases abnormally. The reason behind this is that as the O chemical potential increases, the Fermi level decreases, and the defect density of donor $Ti_{Sr}$ increases as shown in Fig. 4. Especially for +2 charge state $Ti_{Sr}$, its density can go beyond $10^{19}$ cm$^{-3}$. Therefore, $Ti_{Sr}$ is the dominant donor defect under O-rich condition. For $Ti_{Sr}$, it also has a very shallow transition level from +2 to +1 charge state, and it is 0.1 eV below the conduction band minimum (CBM) as shown in Fig. 3(d). In addition, the donor defect $O_{i1}$ also has a relatively low formation energy 0.65 eV for the neutral state, and it has two deep transition levels in the band gap as shown in Fig. 3(d). The first one is from +1 to neutral state (2.85 eV below the CBM), and the second one is from +3 to +1 charge state (3.08 eV below the CBM). These two deep transition levels may serve as carrier trap levels or even recombination centers. Fortunately, the influence of $O_{i1}$ on the conductivity should be small compared with $V_O$ and $Ti_{Sr}$, since it is neutral in most cases when $E_F$ changes in the range of the band gap.

Based on the above discussion, we can determine that the main intrinsic defects are $V_O$, $V_{Sr}$ and $Ti_{Sr}$. This explains why many researchers focus on the vacancies (especially $V_O$) and $Ti_{Sr}$[16,17,25,27,35,36,47,50]. Among all the intrinsic defects, an abnormal phenomenon is observed for $V_{Ti}$, which has a relatively low formation energy under O-rich condition as shown in Fig. 2(c). As we all know, cation vacancy is usually an acceptor, while $V_{Ti}$ is a donor based on our results. In fact, this can be understood after analyzing the defect structure of $V_{Ti}$. As shown in Fig. 3(a), after removing a Ti atom (forming $V_{Ti}$), the O atom (yellow ball) above the removed Ti will break its bond with the Ti atom above it and then move downward, forming an O-trimer. To understand this process clearly, it is illustrated using a planar figure. As shown in Fig. 3(b), one Ti atom (+4 charge state in STO) will bond with four O atoms, forming four single bonds. After removing the left Ti atom as shown in Fig. 3(b), one $V_{Ti}$ and four O dangling bonds will be formed. Usually, four O dangling bonds need four electrons to achieve the full-shell stable state. Hence, $V_{Ti}$ should accept the electron and be an acceptor. However, in STO as shown in Fig.

3(c), the middle O atom (yellow ball) will break its bond with the right Ti atom and form an O-trimer with the left two O atoms. In this situation, the previous four O dangling bonds will change to one O dangling bond and one Ti dangling bond, leaving two unpaired electrons as shown in Fig. 3(c). The unpaired electron of O atom tends to accept one electron, while the unpaired electron of Ti atom tends to donate one electron. Therefore, $V_{Ti}$ could be acceptor, donor or even both. After checking the calculation results, only two single-occupied states (considering the spin polarization) are found in the band gap of STO, and thus $V_{Ti}$ is a donor. For $V_{Ti}$, it has a relatively high formation energy compared with $V_O$ and $V_{Sr}$ (This is consistent with the results reported by Janotti et al.[25]), and hence its influence on the conductivity of STO should be limited.

Fig. 3(d) shows the transition levels of all the intrinsic defects in STO. As we can see, the dominant intrinsic defects either have no transition level in the band gap (like $V_O$), or have very shallow transition levels (like $V_{Sr}$ and $Ti_{Sr}$). This means these defects will never act as the trap states. For the rest intrinsic defects, even though some of them ($Sr_{i2}$, $Ti_{i3}$, $Sr_{Ti}$, $O_{Ti}$ and $V_{Ti}$) have deep transition levels as shown in Fig. 3(d), their influence on the intrinsic carriers and conductivity of STO should be limited, since their formation energies are high, which means their defect densities are low. Meanwhile, most of the intrinsic defects are donor defects. The donor will donate electrons and push up the Fermi level. This could be one of the factors contributing to the unintentional n-type conductivity of STO.

At last, to unravel the origin of oxygen partial pressure-dependent conductivity quantitatively, the density of low-energy intrinsic defects, Fermi level ($E_F$) and intrinsic carrier densities in STO as function of O chemical potential (oxygen partial pressure) are calculated as shown in Fig. 4. Based on the results of $E_F$, the range of O chemical potential can be classified into three typical regions. The first region (R1) is where the $E_F$ is located above the band gap ($E_g$). In this region, $E_F$ goes beyond the CBM and STO shows metallic conductivity when the O chemical potential is poor (oxygen partial pressure is low). At this moment, the density of intrinsic electron carrier is extremely high, close to $10^{20}$ cm$^{-3}$. The dominant intrinsic defects are $V_O^{1+}$, $V_O^{2+}$ and $V_{Sr}^{2-}$, and the density of $V_O^{1+}$ can goes beyond that of $V_O^{2+}$, which can not be directly discovered from the formation energy results as shown in Fig. 2. Moreover, the density of $V_O^0$ can be higher than $10^{23}$ cm$^{-3}$, indicating that STO should be oxygen-deficient under this condition. Intriguingly, the metallic conductivity was reported long ago in oxygen-deficient STO under low oxygen partial pressure[51]. This validates the robustness of our results.

As the increase of oxygen partial pressure, the O chemical potential enters the second region (R2). In this region, STO can show excellent n-type conductivity. In this situation, $E_F$ is very close to CBM (0-0.29 eV below the CBM) and the intrinsic electron carrier density is high ($10^{14}$-$10^{19}$ cm$^{-3}$). At this moment, the donor $V_O^{2+}$, $V_O^{1+}$, $Ti_{Sr}^{2+}$, $Ti_{Sr}^{1+}$ and acceptor $V_{Sr}^{2-}$ are the dominant intrinsic defects affecting the conductivity. Their defect densities are all higher than $10^{15}$ cm$^{-3}$. For the third region (R3), the oxygen partial pressure is very high (O chemical potential is very rich). In this region, the conductivity of STO goes through a transition from n-type to p-type. This can be deduced from the change of the $E_F$ from above the middle of the band gap ($E_g/2$) to below the middle of band gap (about 0.5 eV above the VBM). Meanwhile, the intrinsic carrier changes from electron to hole, and its density is about $10^{11}$ cm$^{-3}$. In this region, STO shows benign p-type conductivity, which is also observed in experiment[52,53]. Furthermore, $O_{i1}^{1+}$ is also an important intrinsic defect in addition to the $Ti_{Sr}^{2+}$, $V_O^{2+}$ and $V_{Sr}^{2-}$ under this condition. Its density is about $10^{15}$-$10^{17}$ cm$^{-3}$.

In summary, the intrinsic defect properties have been systematically studied based on the first-principles calculations. The results show that the dominant intrinsic defects are $V_O$, $V_{Sr}$ and $Ti_{Sr}$. As the change of the O chemical potential, these dominant intrinsic defects can tune the Fermi level from above the middle of the band gap to below the middle of the band gap. Hence, STO can exhibit metallic, excellent n-type and benign p-type conductivity under different O chemical potentials. These findings unravel the origin of the oxygen partial pressure-dependent conductivity from the perspective of intrinsic point defects. Furthermore, our results also uncover the microscopic mechanism behind the experimentally discovered metallic conductivity in oxygen-deficient STO.

**SUPPLEMENTARY MATERIALS**

See the supplementary materials to access the additional details of this work.

**ACKNOWLEDGMENT**

This work was supported by National Natural Science Foundation of China under grant No. 12304110.

## DATA AVAILABILITY

The data that support the findings of this study are available from the corresponding author upon reasonable request.

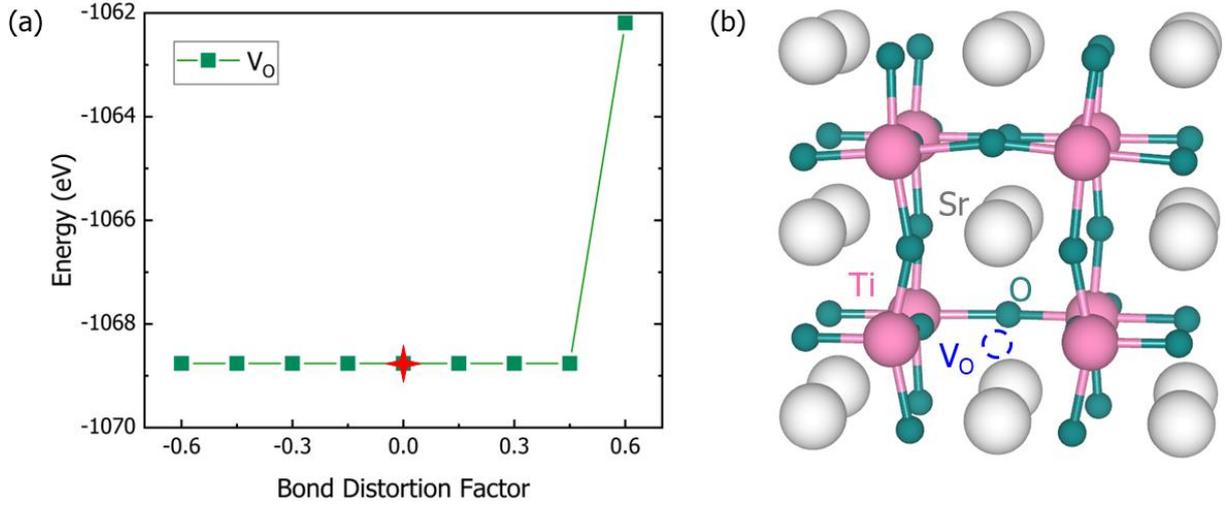

**Figure 1.** (a) Plot of final total energies of neutral $V_O$ versus bond distortion factor. Negative distortion means the adjacent atoms move towards the defect and the positive distortion means the adjacent atoms move away from the defect. Zero means no distortion. All the distortions are relative to the equilibrium lattice position in the perfect cell; (b) The most stable structure of neutral $V_O$ marked with a red star in (a).

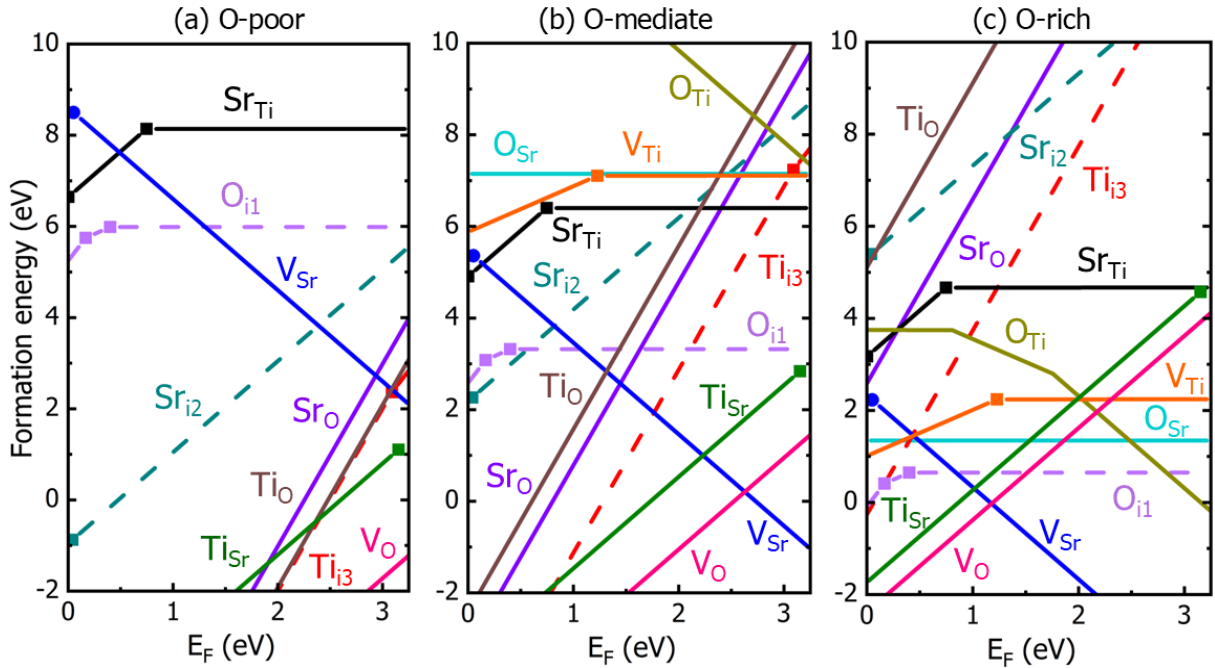

**Figure 2.** The calculated formation energies of intrinsic defects in STO as function of Fermi level ($E_F$) under: (a) O-poor condition ($\mu_O$ = -5.34 eV, $\mu_{Sr}$ = -0.83 eV, $\mu_{Ti}$ = -0.28 eV); (b) O-mediate condition ($\mu_O$ = -2.67 eV, $\mu_{Sr}$ = -3.97 eV, $\mu_{Ti}$ = -5.15 eV; (c) O-rich condition ($\mu_O$ = 0 eV, $\mu_{Sr}$ = -7.10 eV, $\mu_{Ti}$ = -10.02 eV). Square and circle points represent the transition levels of donor and acceptor defects respectively as shown in Fig. 3(d).

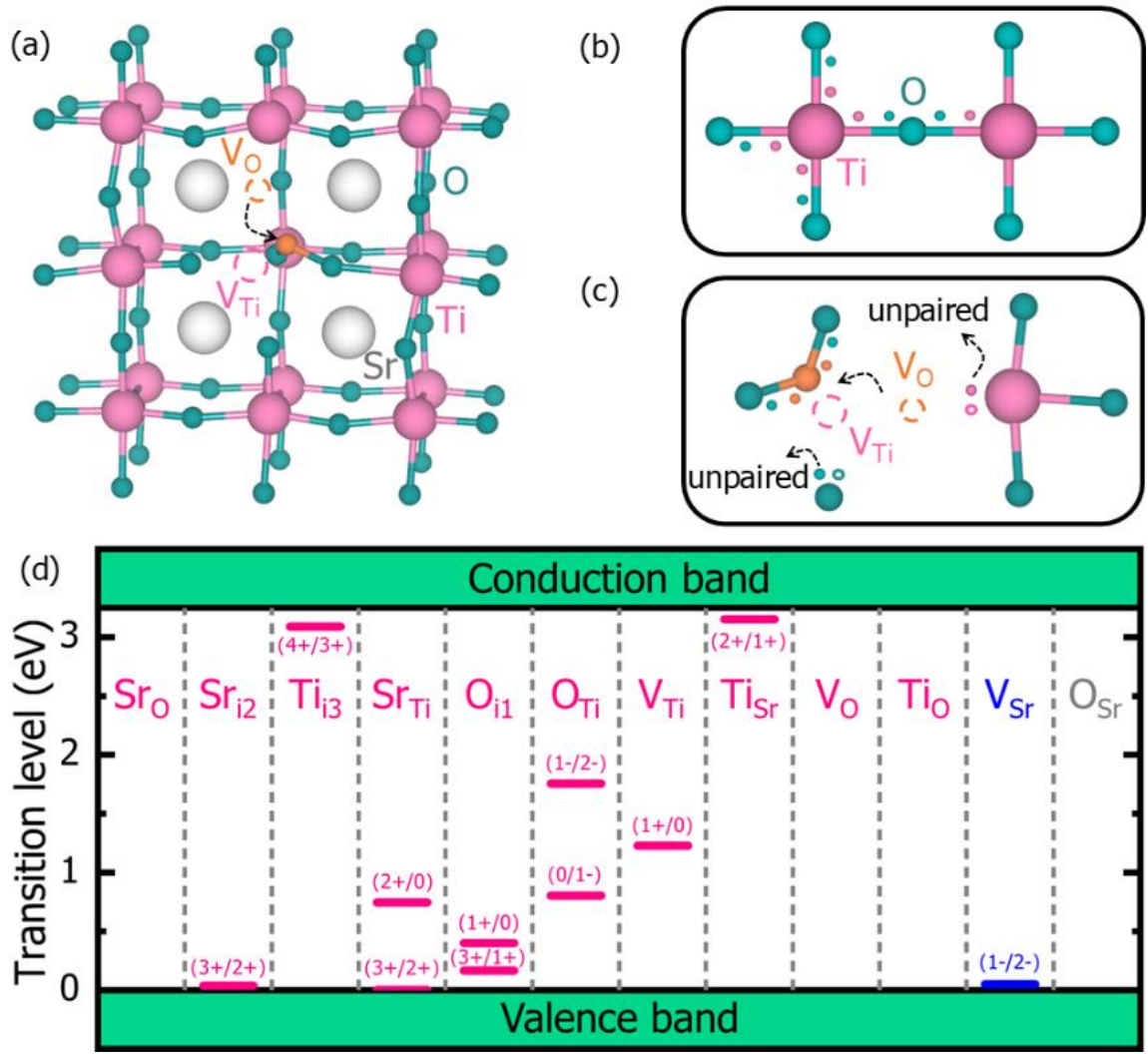

**Figure 3.** (a) The ground-state structure of neutral $V_{Ti}$; (b) and (c) are the bonding situation before and after forming neutral $V_{Ti}$; (d) The calculated transition levels of intrinsic defects in the band gap of STO.

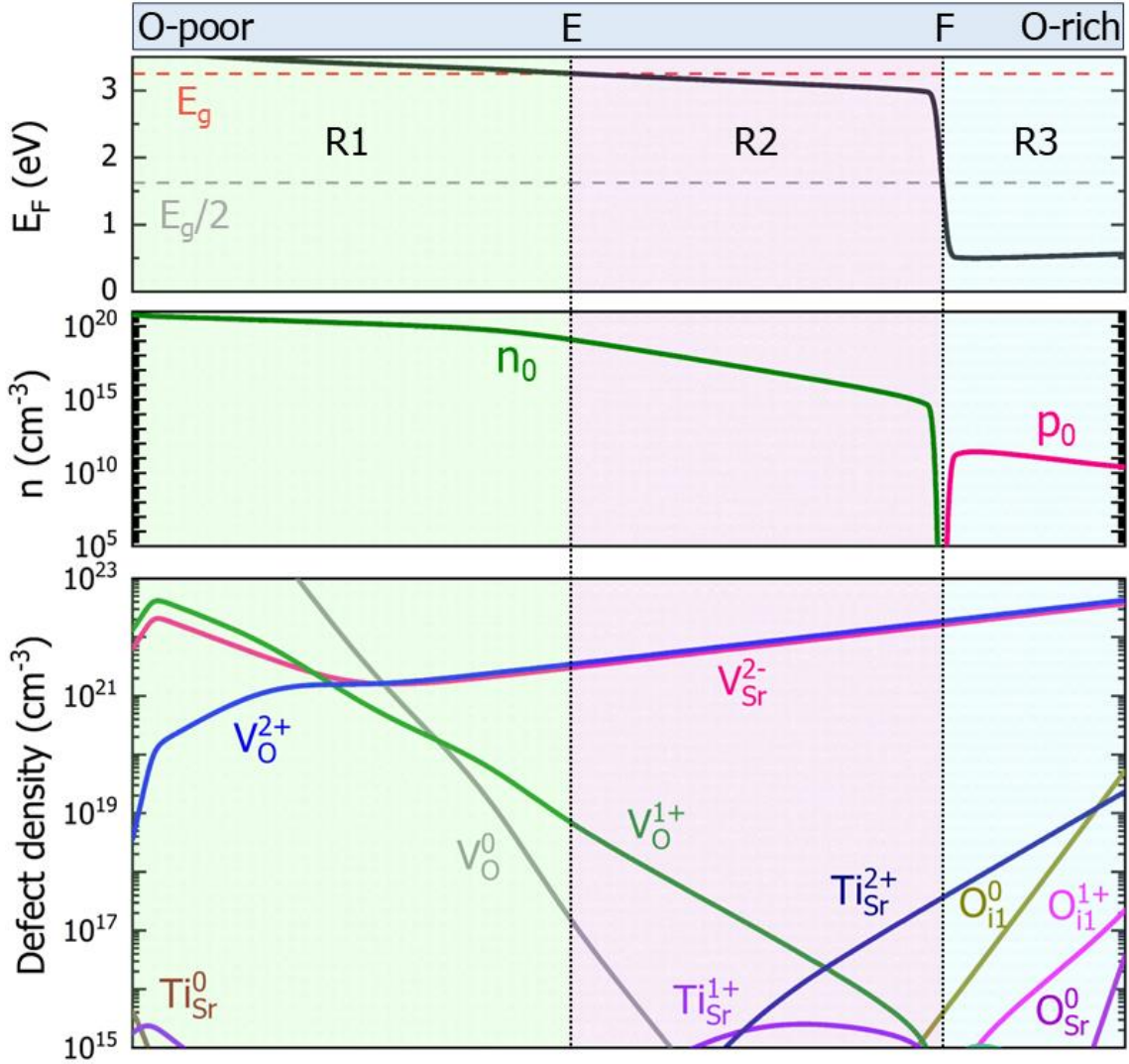

**Figure 4.** The calculated Fermi level ($E_F$), intrinsic carrier densities, and density of low-energy intrinsic defects in STO as function of O chemical potential (The specific chemical potential values of O-rich and O-poor are the same as that in Fig. 2). The middle of the band gap ($E_g/2$) is marked with a grey dashed line. The band gap ($E_g$) is marked with a red dashed line. E ($\mu_O$ = -2.99 eV, $\mu_{Sr}$ = -3.59 eV, $\mu_{Ti}$ = -4.56 eV) is the transition point of conductivity from metallic to n-type. F ($\mu_O$ = -0.99 eV, $\mu_{Sr}$ = -5.95 eV, $\mu_{Ti}$ = -8.22 eV) is the transition point of conductivity from n-type to p-type.